# Integration of SOA and Cloud Computing in RM-ODP

Mostafa Jebbar, Abedrrahim Sekkaki, Othmane Benamar.
Departement of Mathematics and Computer Science
University Hassan II, Ain Chock, Faculty of Sciences
Casablanca, Morocco
mostafajebbar@gmail.com, a.sekkaki@fsac.ac.ma, othmanebenamar@gmail.com

*Abstract*—The objective of ODP is according to ITU-T Recommendation X.901 stated as follows: "The objective of ODP standardization is the development of standards that allow the benefits of distributing information processing services to be realized in an environment of heterogeneous IT resources and multiple organizational domains. These standards address constraints on system specification and the provision of a system infrastructure that accommodate difficulties inherent in the design and programming of distributed systems." This objective seems to cover cloud computing systems. Therefore, we in this paper discuss the concepts of cloud, and discuss the use of RM-ODP for specifying the solution. We indicate that the current RM-ODP may be too abstract for the purpose, and indicate how to adapt RM-ODP to fit the purpose.

## I. Introduction

Businesses have grown in complexity and have become increasingly reliant on information systems. The advent of Cloud Computing technologies opened up opportunities and business evolved different forms such as e-commerce, e-business, supply chains and virtual enterprises. This increased the complexity and challenges for businesses as they struggled to align their IT with their strategic intent. Hence a need arose for a holistic approach to handle this complexity.

In this paper, we look at the concepts of cloud computing. We discuss the capabilities of RM-ODP in solving the complexity and challenges for specifying cloud computing systems. Then we introduce some existing standards in SOA to integrate them into the standard RM-ODP.

## II. Cloud : Definitions and Taxonomy

The following definitions and taxonomy are included to provide an overview of cloud computing concepts.

### A. Definitions of Cloud Computing Concepts Cloud Computing:

Cloud computing is a model for enabling ubiquitous, convenient, on-demand network access to a shared pool of configurable computing resources (e.g., networks, servers, storage, applications, and services) that can be rapidly provisioned and released with minimal management effort or service provider interaction. (This definition is from the latest draft of the NIST Working Definition of Cloud Computing published by the U.S. Government's National Institute of Standards and Technology [2].

#### 1) Delivery Models

The NIST definition of cloud computing defines three delivery models:

- Software as a Service (SaaS).
- Platform as a Service (PaaS).
- Infrastructure as a Service (IaaS).

#### 2) Deployment Models

The NIST definition defines four deployment models:

- Public Cloud.
- Private Cloud.
- Community Cloud.
- Hybrid Cloud.

### B. Taxonomy[1]

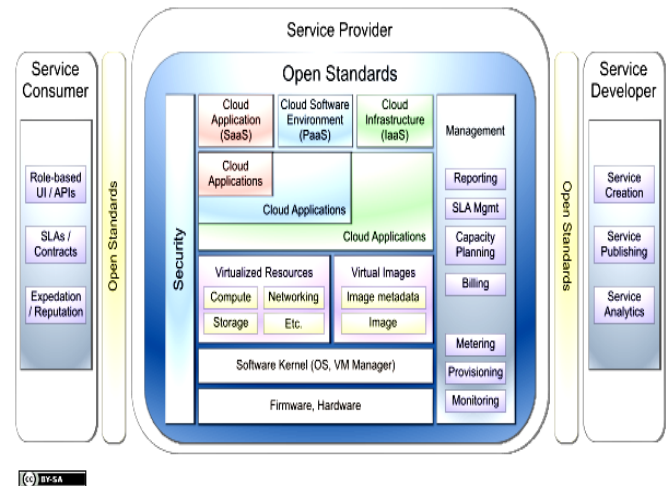

Fig. 1. Diagram defining a taxonomy for cloud computing

In the diagram "Fig. 1", Service Consumers use the services provided through the cloud, Service Providers manage the cloud infrastructure and Service Developers create the services themselves. (Notice that open standards are needed for the interactions between these roles.) Each role is discussed in more detail in [1].

*1) Service Consumer*

The service consumer is the end user or enterprise that actually uses the service, whether it is Software, Platform or Infrastructure as a Service.

Depending on the type of service and their role, the consumer works with different user interfaces and programming interfaces. Some user interfaces look like any other application; the consumer does not need to know about cloud computing as they use the application. Other user interfaces provide administrative functions such as starting and stopping virtual machines or managing cloud storage. Consumers writing application code use different programming interfaces depending on the application they are writing.

Consumers work with SLAs (Service Level Agreements) and contracts as well. Typically these are negotiated via human intervention between the consumer and the provider. The expectations of the consumer and the reputation of the provider are a key part of those negotiations.

*2) Service Provider*

The service provider delivers the service to the consumer. The actual task of the provider varies depending on the type of service

In the service provider diagram, the lowest layer of the stack is the firmware and hardware on which everything else is based. Above that is the software kernel, either the operating system or virtual machine manager that hosts the infrastructure beneath the cloud. The virtualized resources and images include the basic cloud computing services such as processing power, storage and middleware. The virtual images controlled by the VM manager include both the images themselves and the metadata required to manage them.

Crucial to the service provider's operations is the management layer. At a low level, management requires metering to determine who uses the services and to what extent, provisioning to determine how resources are allocated to consumers, and monitoring to track the status of the system and its resources.

At a higher level, management involves billing to recover costs, capacity planning to ensure that consumer demands will be met, SLA management to ensure that the terms of service agreed to by the provider and consumer are adhered to, and reporting for administrators.

Security applies to all aspects of the service provider's operations. Open standards apply to the provider's operations as well. A well-rounded set of standards simplify operations within the provider and interoperability with other providers.

*3) Service Developer*

The service developer creates, publishes and monitors the cloud service. These are typically "line-of-business" applications that are delivered directly to end users via the SaaS model. Applications written at the IaaS and PaaS levels will subsequently be used by SaaS developers and cloud providers.

Development environments for service creation vary. If developers are creating a SaaS application, they are most likely writing code for an environment hosted by a cloud provider. In this case, publishing the service is deploying it to the cloud provider's infrastructure.

During service creation, analytics involve remote debugging to test the service before it is published to consumers. Once the service is published, analytics allow developers to monitor the performance of their service and make changes as necessary.

III. RM-ODP

The Reference Model of Open Distributed Processing RM-ODP[3][4][5][6], is a joint standardization effort by ISO/IEC and ITU-T that creates an architecture within which support of distribution, interworking and portability can be integrated. Several years after its final adoption as ITU-T Recommendation and ISO/IEC International Standard, the RM-ODP is increasingly relevant, mainly because the size and complexity of current IT systems is challenging most of the current software engineering methods and tools. These methods and tools were not conceived for use with large, open and distributed systems, which are precisely the systems that the RM-ODP addresses. In addition, the use of international standards has become the most effective way to achieve the required interoperability between the different parties and the organizations involved in the design and development of complex systems. As a result, we are now witnessing many major companies and organizations investigating RM-ODP as a promising alternative for specifying their IT systems, and for structuring their large-scale distributed software designs especially new generation of Cloud applications.

The RM-ODP provides five "Fig. 2" generic and complementary viewpoints on the system and its environment: enterprise, information, computational, engineering and technology. They allow different participants to observe a system from different perspectives [4]. These viewpoints are sufficiently independent to simplify reasoning about the complete specification of the system. The architecture defined by RM-ODP tries to ensure the mutual consistency among the viewpoints, and the use of a common object model and a common foundation defining concepts used in all of them (composition, type, subtype, actions, etc.) provide the glue that binds them all together.

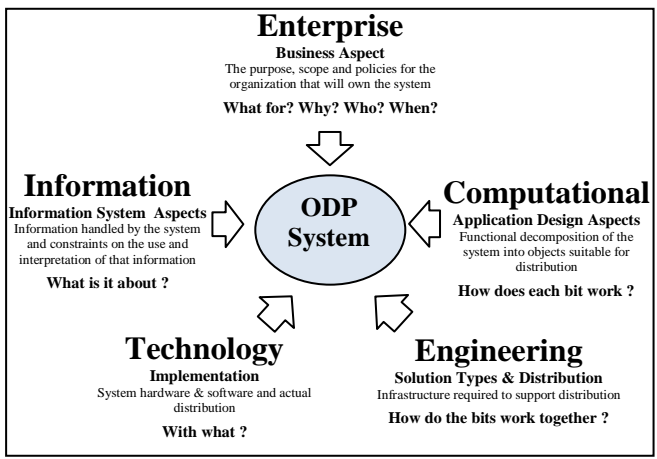

Fig. 2. RM-ODP viewpoints

Although the ODP reference model provides abstract languages for the relevant concepts, it does not prescribe particular notations to be used in the individual viewpoints. UML4ODP [7] defines a set of UML profiles, one for each viewpoint language and one to express the correspondences between viewpoints, by which ODP modelers can use the UML notation for expressing their ODP specifications in a standard graphical way, and UML modelers could use the RM-ODP concepts and mechanisms to structure their large UML system specifications according to a mature and standard proposal.

## IV. CLOUD COMPUTING AND RM-ODP : CONVERGENCE OR DIVERGENCE

By definition Cloud computing is a large distributed system, users in this account per million, working together on computer networks on different applications; The standard RM-ODP aims to large distributed system specification; share this observation we can say that RM-ODP and the cloud converges towards the same objectives (Applications for a large distributed systems and Specification of application of large distributed systems    ). The question is: can RM-ODP support the specification of a cloud applications?

The Reference Model for Open Distributed Processing (RM-ODP) has been published in 1995 and is currently under a revision process that includes better alignment with SOA concepts.

Hence, a new clause "service concepts" has been added to Part 2 (Foundations) [4] and some adaptations have been included in the trading function of Part 3 (Architecture) [5].

For our part we did a study on various SOA standards and Cloud standards which already exists for the proposed extension as the standard RM-ODP.

## V. STANDARDS ANALYSIS UNDER STANDARDS DEVELOPMENT ORGANIZATION

A number of standards consortia are working on the standardization of various aspects of SOA. Many of these standards define terms for SOA, Three of the consortia, The Open Group, OASIS, and OMG, collaborated to understand and position the architectural standards underway. The result of the collaboration is the jointly published positioning paper, "Navigating the SOA Open Standards Landscape Around SOA" [8]

In addition to the standards on architecture, there are standards for implementation and infrastructure for SOA underway in OASIS and OMG as well.

Here we present the standards that can help us make the extension of RM-ODP, The standards organized by the standards organization

### A. Open Group

The Open Group's vision is Boundaryless Information Flow™ which will enable access to integrated information within and between enterprises based on open standards and global interoperability. The Open Group is known for its development of TOGAF and Unix.

#### 1) SOA Reference Architecture

The draft standard SOA reference architecture [12] uses a partially layered approach since one layer does not solely depend upon the adjacent layers. Layers are defined around sets of key architectural concerns and capabilities, the interaction protocols between layers, and the details within a layer using a set of architectural building blocks. There are five functional horizontal layers and four non-functional vertical layers that support various cross-cutting concerns of the SOA architectural style.

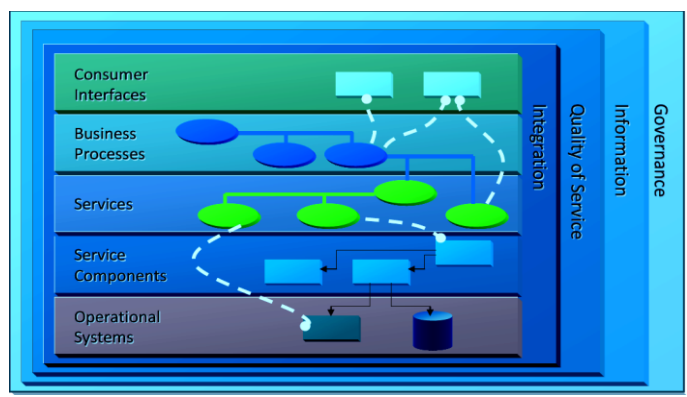

Fig. 3. Layers of the SOA Reference Architecture

Three of the layers "Fig. 3" address the implementation and interface with a service (the Operational Systems Layer, the Service Components Layer and the Services Layer). Two of them support the consumption of services (the Business Process Layer and the Consumer Layer). Four of them support cross-cutting concerns of a more "non-functional" nature (the Information Architecture, Quality of Service, Integration and Governance Layers). The RA-SOA as a whole provides the framework for the support of all the elements of a SOA, including all the components that support services, and their interactions.

This logical view of the SOA Reference Architecture addresses the question, "If I build a SOA, what would it look like and what abstractions should be present?"

The SOA Reference Architecture enumerates the fundamental elements of a SOA solution and provides the architectural foundation for the solution..

*2) Service Oriented Cloud Computing Infrastructure (SOCCI)*

The goal of The Service-Oriented Cloud Computing Infrastructure [13] project will provide recommendations and guidelines that enable the provisioning of infrastructure as a service in the SOA solutions and cloud computing environments.

This draft will cover:
1. Definition of Service Oriented Cloud Computing Infrastructure, SOI and Infrastructure as a Service (IaaS)
2. Identify required components for enabling Service-Oriented Infrastructure as a Cloud Service and SOA service
3. Application of Enterprise Service Management concepts
4. Define relationship between SOA and XaaS (Business Process (BPaaS), Software (SaaS), Platform (PaaS), and Infrastructure (IaaS))
5. Define consumption models for IaaS

This specification is important to SOA standards as it defines how to expose existing hardware, infrastructure, software, and virtualized versions of these as services that can be used equally by SOA solutions and Cloud computing.

*B. OASIS*

OASIS (Organization for the Advancement of Structured Information Standards) is well known for Web services standards along with standards for security, e-business, the public sector and application-specific markets

*1) Reference Model for Service Oriented Architecture 1.0*

The reference model [10] provides a normative reference that remains relevant for SOA as an abstract, powerful model, regardless of the inevitable technology changes that have influenced or will influence SOA deployment. The SOA RM is an abstract framework for understanding significant entities and relationships between them within a service-oriented environment, and for the development of consistent standards or specifications supporting that environment. It is based on unifying concepts of SOA and may be used by architects developing specific service oriented architectures or in training and explaining SOA.

The OASIS SOA Reference Model "Fig. 4" applies directly to the Vocabulary category of RM-ODP.

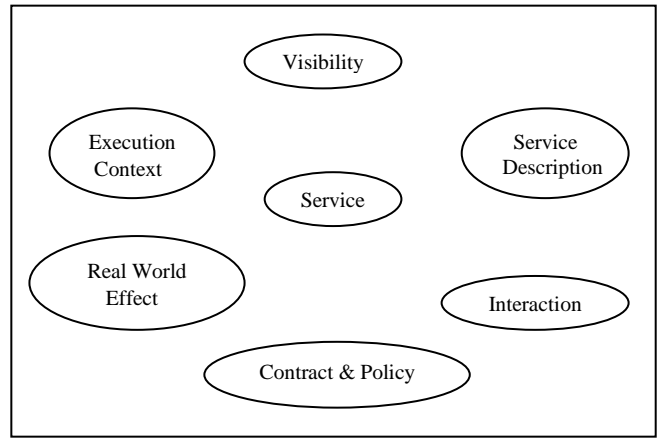

Fig. 4. Principal concepts in the OASIS SOA Reference Model

*2) OASIS Reference Architecture for SOA Foundation Version 1.1*

The goal of the OASIS Reference Architecture for SOA Foundation [9] is to define a view-based abstract reference architecture foundation that models SOA from an ecosystem/paradigm perspective. It specifies three viewpoints; specifically, the *Service Ecosystem* viewpoint, the *Realizing SOAs* viewpoint, and the *Owning SOAs* viewpoint. It is based on the concepts and relationships defined in the OASIS Reference Model for Service Oriented Architecture. Each of the associated views that are obtained from these three viewpoints is briefly described below.

The *Service Ecosystem* view "Fig. 5" contains models that are intended to capture how SOA integrates with and supports the service model from the perspective of the people who perform their tasks and achieve their goals as mediated by SOAs. Since the Service Ecosystem viewpoint (on which this view is based) emphasizes the use of SOA to allow people to access and provide services that cross ownership boundaries, it is explicit about those boundaries and what it means to cross an ownership boundary.

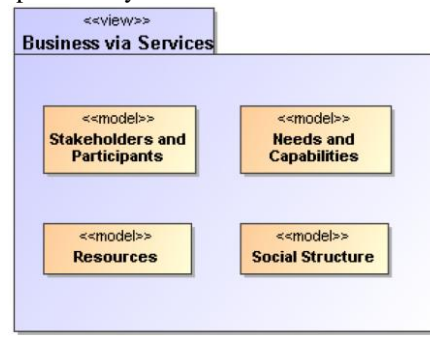

Fig. 5. Model elements described in the Business via Services view

The *Realizing SOAs* view "Fig. 6" contains models for description of, visibility of, interaction with, and policies for services.

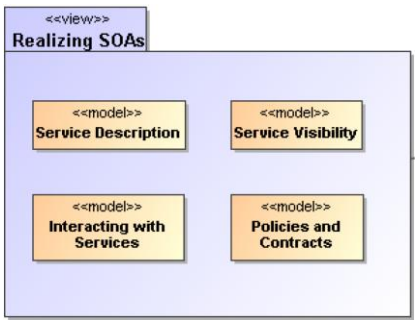

Fig. 6. Model Elements Described in the Realizing a Service Oriented Architecture View

The *Owning SOAs* view "Fig. 7" contains models for securing, managing, governing, and testing SOA-based systems.

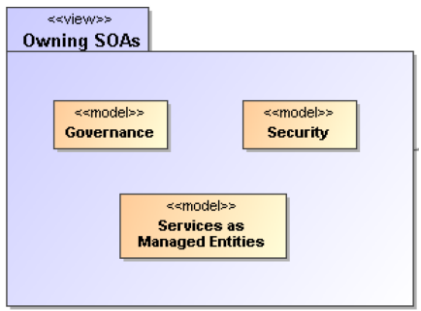

Fig. 7. Model elements described in the Owning Service Oriented Architectures view

This Reference Architecture is principally targeted at Enterprise Architects; however, Business and IT Architects as well as CIOs and other senior executives involved in strategic business and IT planning should also find the architectural views and models described to be of value.

*C. OMG*

OMG has been an international, open membership, not-for-profit computer industry consortium since 1989. OMG Task Forces develop enterprise integration standards for a wide range of technologies. OMG's modeling standards, including the Unified Modeling Language™ (UML®) and Model Driven Architecture® (MDA®), enable powerful visual design, execution and maintenance of software and other processes, including IT Systems Modeling and Business Process Management.

*1) Service oriented architecture Modeling Language (SoaML)*

The goals of SoaML [11] are to define extensions to UML for services modeling and provide functional, component, and service-oriented modeling capabilities. SoaML extends UML in order to provide additional capabilities for managing cohesion and coupling afforded by an SOA style. The standard is intended to be sufficiently detailed to define platform-independent SOA models (PIM) that can be transformed into platform-specific models (PSM) for particular technical architectures as described by the OMG MDA. The intent of SoaML was to provide a foundation for integration, interoperability, and extension.

The fundamental element of SoaML "Fig. 8" is the participant, representing a service consumer and/or provider. Participants express their goals, needs, and expectations through requests for services as defined by service interfaces or service contracts. Other participants express their value propositions, capabilities, and commitments through services. Participants are then assembled into service value chains where participant requests are connected to the compatible services of other participants through service channels through which they interact. SoaML uses facilities of UML to define the services interfaces and method behaviours for carrying out and using services. SoaML also defines autonomous agents that can choreograph participants in a service value chain while adapting to the changing needs of the community of collaborating participants. SoaML provides a means of defining milestones that indicate the achievement of progress toward achieving the desired real-world effect of the services value chain, and for evaluating different approaches to achieving progress by different participants.

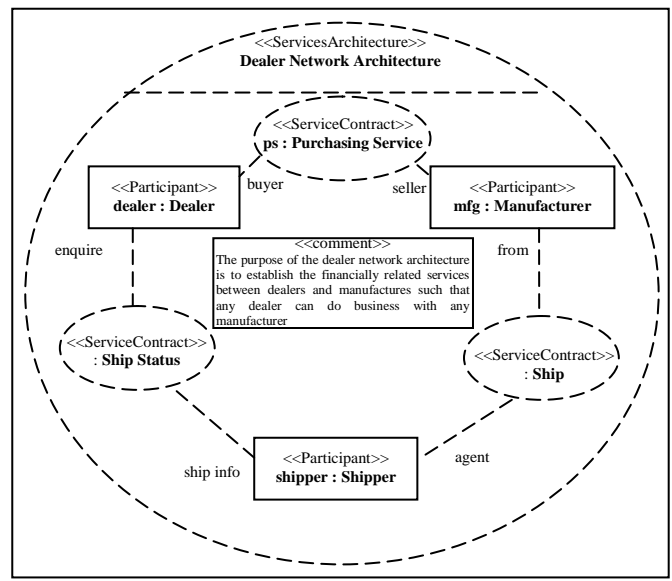

Fig. 8. Example community services architecture with participant roles and services

## VI. EXTENTIONS OF RM-ODP

From the previous paragraph it seems that the different standards are in line with RM-ODP, and to integrate the SOA into RM-ODP, we can take different concepts existing in these standards and reused in RM-ODP.

### A. Integration of OASIS SOA-RM in RM-ODP Foundation (ISO / IEC 10746-2):

The Reference Model for Service Oriented Architecture is intended to capture the "essence" of SOA, as well as provide a vocabulary and common understanding of SOA. The goals of the reference model include a common conceptual framework that can be used consistently across and between different SOA implementations, common semantics that can be used unambiguously in modeling specific SOA solutions, unifying concepts to explain and underpin a generic design template supporting a specific SOA, and definitions that should apply to all SOA.

While Rec. ITU-T X.902 | ISO / IEC 10746-2 Foundations: it contains the definition of concepts and analytical framework to be used for the standardized description of distributed processing systems (arbitrary). She sticks to a level of detail sufficient to support Rec. ITU-T X.903 | ISO / IEC 10746-3 and establish requirements for new specification techniques.

It is noted that OASIS RM-SOA are in the same direction, so we take all the necessary definitions to SOA from SOA-RM and joined it in (Concepts services) of ISO / IEC 10746-2.

### B. Integration of RA-SOA of the Open Group and OASIS SOA-RA in RM-ODP Architecture (ISO / IEC 10746-3)

The SOA reference architecture of the Open Group uses a layered approach as a partial layer does not only depend on adjacent layers. The layers are defined around a set of architectural concepts, protocols of interaction between the layers and details in a layer with a set of architectural building blocks. There are five functional layers and four horizontal layers non-functional vertical support issues cross the style of SOA.

The reference architecture of OASIS SOA specifies three points of view, more precisely, the view through business service point of view of achievements SOA, and SOA perspective of the owner. It is based on the concepts and relationships defined in the model reference for the OASIS SOA. Each of the points of view represented in the form of UML diagram.

On the other hand Rec. ITU-T X.903 | ISO / IEC 10746-3: Architecture: it decomposes a system into five Viewpoint, business, information, computational, engineering, technology.

It is found that RA-SOA from the Open Group and OASIS RA-SOA and RM-ODP Architecture (ISO / IEC 10746-3) are along the same lines, The Open Group RA-SOA decompose the system in five functional layers, four no-functional layers, all this is a point of view soa, while OASIS SOA-RA decompose the system in three views to a point of view soa, while RM-ODP Architecture (ISO / IEC 10746-3) decomposes the system in 5 viewpoints that includes all the system.

From our point of view, the same concepts of decomposition to view that suggest the two RA-SOA are already existing in RM-ODP architecture, it is sufficient to include in each viewpoint of RM-ODP Architecture a sub view SOA, These subs views take their concepts from RA-SOA OASIS and RA-SOA OPEN GROUP.

### C. INTEGRATION OF SOAML IN UML4ODP:

SoaML objectives are to define extensions to UML for modeling services and provide functional components and how service-oriented modeling. Each of these modeling approaches provide different and improved ways to deal with cohesion and coupling in complex systems. SoaML extends UML to provide additional features to manage cohesion and coupling offered by an SOA style.

On the other hand UML4ODP fact the extension of UML for modeling of the five viewpoints RM-ODP, and the correspondence between views, and sets for each language point of view, a meta model. With the adoption of SOA by enterprises and the emergence of cloud computing, new needs appear and the specification of the new system based on SOA and cloud remains impossible, because the concept of service itself is not defined in RM-ODP.

SoaML support the instantiation of the SOA reference model OASIS [11] and provides a concrete platform for modeling integrated with UML and supporting the OMG's MDA. This profile modeling can be used in conjunction with the normalization in the standard ODP precisely UML4ODP. Use language common to these various modeling systems and the integration of separate fields to enable business agility that can be represented by models of architectural art.

We propose to take the new profiles UML defines SOAML and integrated into the UML4ODP to enable interoperability with already existing, and secondly to allow the modeling of system specifications based on SOA and Cloud Computing.

## VII. EXAMPLE USE CASE

In contrast to most LBSs for tourism, which represents the user geo-referenced information about tourist attractions and other POIs, the presented application [14] tries to give the user a ranked list of tourist attractions. The ranking is calculated using a MCE, based on user preferences, e.g., personal interests and the actual location. The following scenario shall be solved efficiently with such a system: A tourist visits a city, where she or he is not familiar with. The tourist has huge interest in the culture and architecture of the city and moderate interest in nature and parks as well as shopping and events. She/he wants to see places other people recommend or many people think that these are worth visiting. The tourist wants to have a suggestion of attractive places, and how well they fit to her/his preferences. Based on such a list the tourist wants to make a decision which places are worth to visit for her/him and select these tourist attractions. Additional she/he needs to receive relevant information and the routing to the selected places.

### A. Enterprise viewpoint and requirements

With the enterprise viewpoint tries to specify the scope of the tourist guide application. Figure 9 emphasises on the purpose and scope of the system and specifies which actors and use cases are involved in the system process. The diagram tries to show the roles of the actors and which basic activities have to be performed for the objective of the application

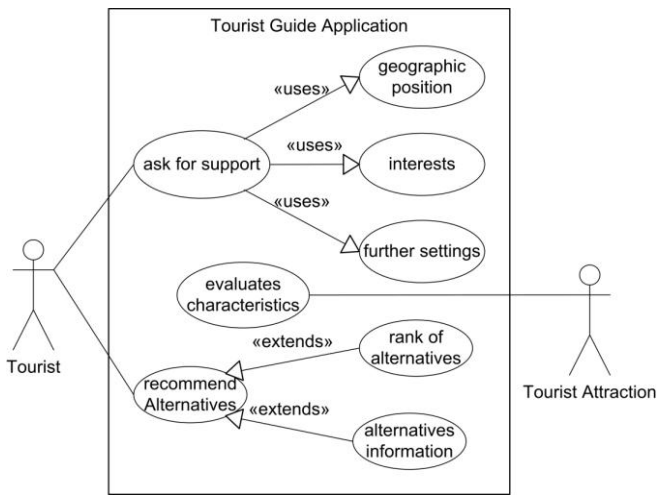

Fig. 9. Use Case diagram of the tourist guide application.

The *Use Case* diagram of Figure 9 identifies actors like *Tourist* and *Tourist Attraction*. The Tourist has the role of the consumer of tourist guide application and wants to visit some tourist attractions within a city. Therefore the user asks the tourist guide for support. The tourist guide evaluates the characteristics of the tourist attractions against the tourist interests, the position and further settings to recommend a list of ranked attractions to the tourist.

### B. Information management and processing

The information viewpoint is concerned with information modelling. An information specification defines the semantics of information and the semantics of information processing in an ODP system, without considerations about other system details, such as its implementation, or the technology used to implement the system.

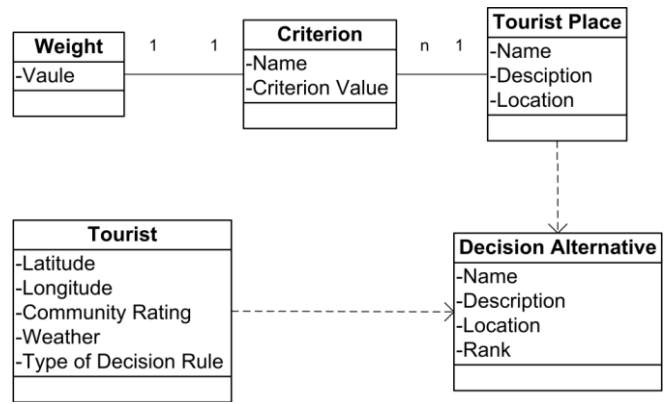

Fig. 10. Static schema of exchanged information.

The selection of evaluation criteria and data sources for tourist places plays a role to understand the information management and processing.

### C. Computational Viewpoint

The computational viewpoint describes the functionality of the tourist guide LBDS application and its environment through the decomposition of the system, in distributed transparent terms, into objects which interact at communication interfaces. In the computational viewpoint, application and distributed functions consist of configurations of interacting computational objects. The computational viewpoint is directly concerned with the distribution of processing but not with the interaction mechanisms that enable distribution to occur. The computational specification decomposes the system into objects performing individual functions and interacting at well-defined interfaces.

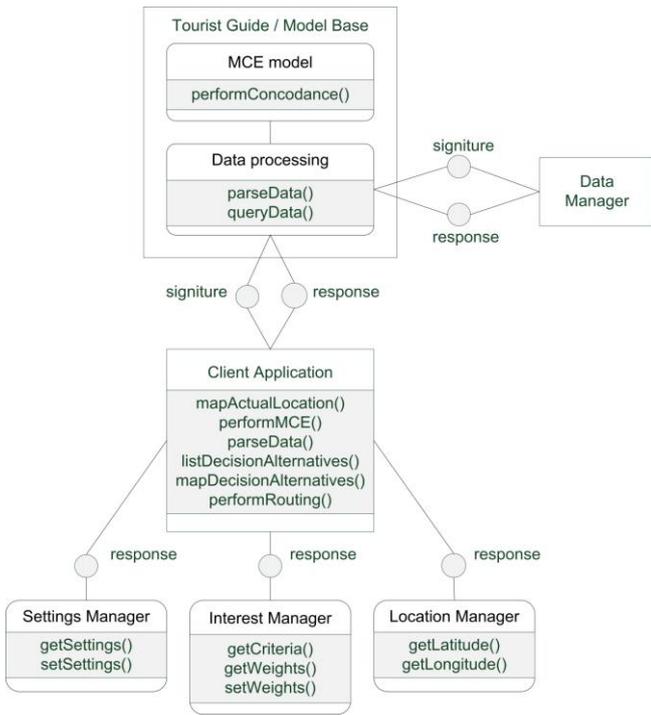

Fig. 11. Computational viewpoint of the tourist guide application, including functional objects and important operations.

### D. Engineering perspective

The engineering viewpoint focuses on the mechanisms and functions required to support distributed interactions between objects in the system. It describes the distribution of processing performed by the system to manage the information and provide the functionality.

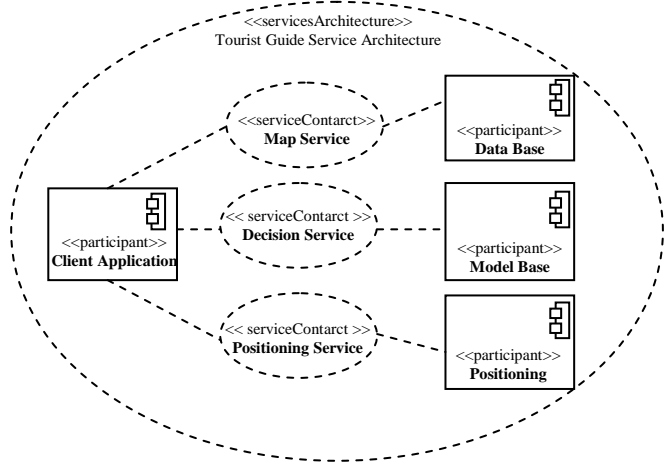

Fig. 12 . Engineering viewpoint of the tourist guide application, we use SoaML modelisation

### E. Technological perspective

The technological perspective should give the idea which real-world software, hardware and network components are used. This is the starting point for the engineering process. Figure 12 shows the technological viewpoint of the LBDS application including their components. The data layer consists of content relevant geographic data, which are in the prototype application information about the tourist attractions to form decision alternatives. Additional data like community data is coming from third-party web services. The logic layer includes elements for data merging and conversion and MCE techniques. In the tourist guide application the concordance method is used as decision rule. The communication between data layer and logic layer is based on the Internet. Also the communication between the mobile client and the logic layer is also done via the Internet using SOAP web services. The client includes a part for the management and communication with the model base, the UI and positioning technology.

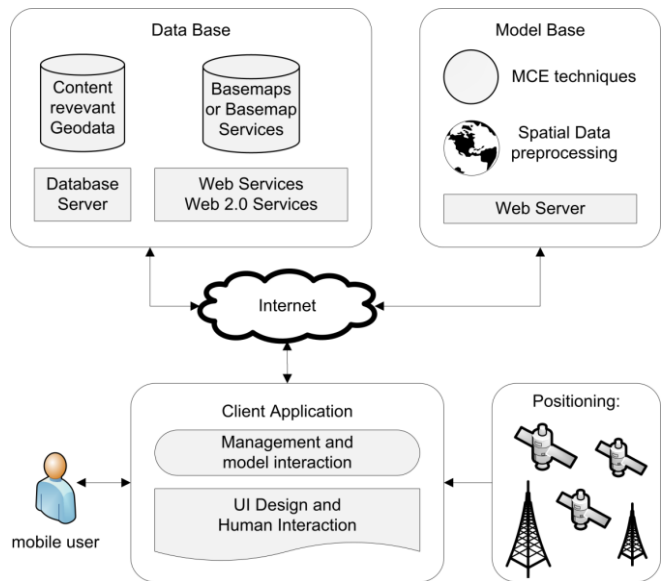

Fig. 12. LBDS architecture from the technology view.

### VIII. CONCLUSIONS AND FUTURE WORK

Our study focused on the development of RM-ODP, after a literature review on the field SOA and Cloud Computing, we found the following possible changes in RM-ODP
- RM-ODP does not incorporate the concept of SOA
- RM-ODP does not incorporate the notion of Cloud Computing

Building on these, we proceeded to identify the various SOA standards developed or being developed by different international consortia; and we make a comparison with the concepts of RM-ODP, and see what is possible to integrate with it.

We found four existing SOA standards that can be integrated with the RM-ODP, SOA-RM of the Open Group RA-SOA Open Group, SOA-RA OASIS, and SOAML the OMG.

Our future work will focus on the integration of new concepts of these standards and we will give the various topics that must be integrated or update in RM-ODP.